\documentclass[prc,preprint]{revtex4}

\usepackage{graphicx}
\usepackage{amssymb}

\begin{document}
\title{\bf The structure of cold neutron star with a quark core within the MIT and NJL models}

\author{{ T. Yazdizadeh$^{1}$
\footnote{E-mail:
tyazdizadeh@yahoo.comr}},{G. H. Bordbar $^{2, 3, 4}$\footnote{Corresponding Author. email: ghbordbar@shirazu.ac.ir}}}

\affiliation{$^1$Department of Physics, Payame Noor University (PNU), P.O. Box 19395-3697, Tehran, Iran\\
$^2$Department of Physics,
Shiraz University, Shiraz 71454, Iran\footnote{Permanent address}\\
$^3$Research Institute for Astronomy and Astrophysics of Maragha,
PO Box 55134-441, Maragha, Iran\\
$^4$ Department of Physics and Astronomy, University of Waterloo,
200 University Avenue West, Waterloo, Ontario N2L3G1,
Canada}
%%%%%%%%%%%%%%%%%%%%%%%%%%%%%%%%%%%%%%%%%%%%%%%%%%%%%%%%%%%%%%%%%%%%%%%%

\begin{abstract}
Neutron star due to their high interior matter density are expected to be
composed of a quark core, a mixed quark-hadron  matter, and a layer
of hadronic matter. Thus, in this paper, we compute the equation
of state of these parts of neutron star to evaluate its
structure properties. We use two  models for describing EOS of quark matter, NJL and MIT bag models, and
employ three approaches in this work. A density dependent bag
constant satisfy the quark confinement in the simple MIT bag model. We also
study the interaction behavior of quarks, firstly one
gluon exchange within MIT bag model and the secondly dynamical mass will be held as
effective interaction that roles between particles. Density
dependence of quark mass is obtained from NJL self consistent model.
NJL model is a effective manner for justify the chiral symmetry.
Applying the Gibbs conditions the equation of state of the quarks
and hadrons mixed phase is obtained. Since the hadronic matter is
under the influence of strong force of nucleons, we calculate the
equation of state of this phase using a powerful variational
many-body technique. Finally, we calculate the mass and radius
of a cold neutron star with a quark core by numerically solving the
TOV equation. To check our used EOS, we compare our results with the
recent observational data. Our results are in a good agreement with some observed compact objects such as  $SAXJ1748.9-2021$, $4U1608-52$  and $Vela X-1$.
\end{abstract}
\pacs{21.65.-f, 26.60.-c, 64.70.-p}
%%%%%%%%%%%%%%%%%%%%%%%%%%%%%%%%%%%%%%%%%%%%%%%%%%%%%%%
\maketitle

%%%%%%%%%%%%%%%%%%%%%%%%%%%%%%%%%%%%%%%%%%%%%%%%%%%%%%%%%%%%%%%%%%%%
\noindent {\bf Keywords:} neutron star, quark core, quark matter,
MIT bag model, NJL model, gravitational mass, radius.

\section{INTRODUCTION}
Neutron stars are placed in category of compact objects, and the
interior matter of them can reach a density much greater than the
normal nuclear saturation density. Therefore, these astrophysical objects are best
laboratory and a unique environment  to probe the properties of dense matter. Studying these stars is one of main
problems in physics. In high densities, hadrons
dissolve to quarks, and a phase transition is happen from hadronic
matter to quark matter. Many years ago, the presence of quark matter
in neutron stars has been suggested by  Ivanenko\cite{1}, Itoh \cite{2} and Collins \cite{3}.
 There are up, down, and strange  quarks in the  quark matter, and this strange
matter is a fermi gas which the other quarks because of their high
masses do not appear in this part. Since all the hadrons do not
converge to quarks simultaneously, it is expected to exist a mixed
phase of quarks and hadrons at finite range of density that the
energy is lower than that of quark and hadron matters.

Historically Glendenning was the first who pointed to the neutrality
charge of two phases in mixed phase \cite{4}. In mixed
phase, we study the transition from a hadron phase to quark
phase using the Gibbs conditions. Since the existence of mixed
phase of quarks and hadrons affects the properties of neutron
star, we consider the neutron stars to be composed of a quark matter core,
a mixed phase of quarks and hadrons and a layer of hadrons.
There is a high uncertainty in equation of state (EOS) of quark matter.
Usually, two more efficient models are used to study
 deconfined quark matter, the MIT bag model
\cite{5, 6} and the Nambu-Jona-Lasinio (NJL)
model \cite{7}. The total energy density in MIT bag model
is the sum of the kinetic energy of free quarks and a bag constant
$B$ that is nonperturbative energy shift. We will be search influence
of one-gluon-exchange in MIT bag model. There is  asymptotically
interaction among quarks at high densities. This interaction can show
up by  one gluon exchange. Therefore, in addition of $B$, we add
another term to EOS that is identified by $\alpha$ the QCD coupling
constant \cite{8, 9, 10}. In MIT bag model,
the mass of quarks is constant, but in the second method, NJL, mass
of quarks  depends on density that is considered  as the effective
interaction of quarks. Historically first time the NJL model is
presented in two papers  from Nambu and Jona-Lasinio in 1961
\cite{11, 12}.
 In spite of MIT bag model, the
NJL model does not have confinement, but it justifies chiral symmetry.
But at high density matter like quark matter,  although the confinement
and chiral symmetry are in the least importance, both models treat to be similar \cite{13}.

In recent years, we have applied the MIT bag model to investigate
the cold and hot strange quark star. For example we consider MIT bag model with a density dependent bag constant for a hot strange star  in \cite{14}, and found that the mass and radius of strange star decreases when temperature increases. We have also found that a higher mass and radius will be obtained for a density dependent bag constant compare to a fixed bag constant. In \cite{15,16} using MIT bag model with fixed bag constant, we considered the stability of spin polarized quark star in a strong magnetic field compared to unpolarized case, and calculated the structure of this star at zero and finite temperatures. Also We have used MIT bag model with a density dependent bag constant to calculate the structure of spin polarized strange star in presence of magnetic field at zero and finite temperatures \cite{17,18}. We have also applied NJL model to calculate the equation of state of quark matter
\cite{19}.

As we know, a neutron star with a quark core is called hybrid star. In our previous works, we have obtained the
structure of hybrid star by MIT bag model with fixed and density dependent bag constant
model at zero temperature \cite{20} and at finite temperature
\cite{ 21, 22}. In those works, we have considered the simplest version of the MIT bag model. In the present work, we
intend to develop our previous calculations by considering the effects
of one gluon exchange for the quark matter in a neutron star with a quark core. We also use the NJL method  in this work.
NJL model is an effective theory and a good choice for the studying the chiral quark and diquark condensates.
Chiral symmetry and its breakdown in vacuum are the basic property of NJL model.
The outline of our work is as follows: In section \ref{Energy
calculation}, we calculate the equation of state  of three mentioned
phases of the neutron star matter. Then using this equation of state, we determine
the mass and radius of neutron star with a quark
core in section \ref{structure}.
%-----------------------------------------------------------

\section{equation of state of a hybrid neutron star}
\label{Energy calculation}
 Here, we determine EOS of different part of neutron star: a hadron
phase, quark phase and a mixed phase of quarks and hadrons
respectively.

%%%%%%%%%%%%%%%%%%%%%%%%%%%%%%%%%%%%%%%%%%%%%%%%%%%%%%%%%%%%%%%%%%%%%%%%%%%%%%%%%%%%%%%%%%%%%%

\subsection{Hadron Phase}
We consider the lowest order constrained variational (LOCV)
many-body method for hadron phase \cite{23, 24, 25, 26}.
%\cite{23, 24, 25, 26, 27, 28, 29, 30}.
%
By considering a many body trail wave function such as $\psi=F\phi$
and some calculations, the cluster expansion of the energy functional
is gotten \cite{31},
\begin{eqnarray}\label{eq3}
E([f])&=&\frac{1}{A}\frac{<\psi\mid H \mid\ \psi>}{<\psi\mid
\psi>}\nonumber\\&=&E_1+E_2+E_3+\cdots,
\end{eqnarray}
where $F=S\prod_{i<j} f(ij)$ is an A-body correlation operator
($f(ij)$ is two-body correlation function and $S$ leads to a symmetric product)
and $\phi$ is the slater determinate of $A$ noninteracting nucleons.
We consider the first two terms in above equation, the one-body term
$E_1=\sum_{i=1,2}{\frac{3}{5}\frac{k_i^2}{2m_i}\frac{\rho_i}{\rho}}$, and
the two-body
term $E_2=\frac{1}{2A}\sum_{ij}<ij|\nu(12)|ij-ji>$.
In these relations, $\rho_i$ is the nucleon density,
$\rho=\rho_p+\rho_n$ is the total density and $k_i=(3\pi^2\rho_i)^{1/3}$.
Here $\nu = -\frac{\hbar^2}{2m}[f(12),[{\nabla_{12}}^2,f(12)]]+f(12)V(12)f(12)$ is the nucleonic effective potential ($V(12)$ is the nuclear potential).  See reference
\cite{24} for full nuclear matter calculations.

%%%%%%%%%%%%%%%%%%%%%%%%%%%%%%%%%%%%%%%%%%%%%%%%%%%%%%%%%%%%%%%%%%%%%%%%%%%%%%%%%%%%%%%%%%%%%%%%%%

\subsection{Quark Phase}
 We employ MIT bag
model and NJL model which are two well-known and efficient models
for describing the characteristics of deconfined quark matter.
 \subsubsection{The MIT Bag Model}
 Quark matter is a fermi gas which composed of deconfined up, down,
 and strange quarks. Therefore the total energy is given by
 \begin{equation}\label{eq8}
    {\cal E}_{tot} = {\cal E}_u + {\cal E}_d + {\cal E}_s + B.
\end{equation}
In equation (\ref{eq8})  ${\cal E}_i$ is
\begin{eqnarray}\label{eq9}
    {\cal
    E}_i&=&\frac{3m_i^4}{8\pi^2}\left[x_i(2x_i^2+1)(\sqrt{1+x_i^2})-\sinh
    ^{-1} x_i\right]\nonumber\\&&-\alpha_c\frac{m_i^4}{\pi^3}\left[x_i^4-\frac{3}{2}[x_i(\sqrt{1+x_i^2})-\sinh
    ^{-1} x_i]^2\right],
\end{eqnarray}
where
\begin{equation}\label{eq10}
 x_i=\frac{k_F^{(i)}}{m_i}.
\end{equation}
Here, $k_F^{(i)}=(\rho_i\pi^2)^{1/3}$ , $m_i$ and  $\rho_i$ are the mass and baryon density of quark $i$, respectively, and $\alpha_c$ is the QCD
coupling constant. We consider three values for $\alpha$: $\alpha=0
$ (MIT bag model without interaction), $\alpha=0.16$ and $\alpha=0.5$. Although these  amount of $\alpha$ are small and perturbative, they show an appropriate range of  quark interaction and
 are in the selection rang of Farhi and Jaffe work \cite{f}.
 In equation (\ref{eq8}), $B$  is a density
dependent bag constant which satisfies the quark confinement in MIT
bag model. We consider a Gaussian form
$B(\rho)=B_\infty+(B_0-B_\infty)\exp\left[{-\beta(\frac{\rho}{\rho_0})^2}\right]$,
where $B_0=B(\rho=0)=400MeV/fm^3$, $\beta$ is a numerical parameter
equal to $\rho_0=0.17fm^{-3}$ and $B_\infty$ is a free parameter which is determined by using the experimental data reported in the CERN SPS.
\cite{20, 34}.
 Now, using the energy density from
Eq. (\ref{eq8}), the EOS of quark matter in the MIT bag model is obtained,
\begin{equation}\label{eq12}
P(\rho)=\rho\frac{\partial{\cal E}}{\partial\rho}-{\cal E}.
\end{equation}

\subsubsection{The NJL Model}
In NJL model, The dynamical mass is held the effective interaction
between particles. In this method, we adopt a lagrangian similar to that given in reference
\cite{19}, as follow,
% common three flavor
%which preserves chiral symmetry of QCD
\begin{eqnarray}\label{eq13}
{\cal L}&=&\bar{q}\left(i\gamma^\mu\partial_\mu-\hat{m}_0\right)q+
G\sum_{k=0}^8\left[(\bar{q}\lambda_kq)^2+(\bar{q}i\gamma_5\lambda_kq)^2\right]\nonumber\\&&-
K\left[det_f(\bar{q}(1+\gamma_5)q)+det_f(\bar{q}(1-\gamma_5)q)\right] ,
\end{eqnarray}
where $q$ is the  field of quarks with three flavors and three
colors,  $\hat{m}_0=diag(m_0^u,m_0^d,m_0^s)$ in flavor space, and
$\lambda_k$ $(0\leq k\leq8)$ are the  flavor matrices. Restoring
chiral symmetry, breaking is indicated with a ultra-violet cut-off.
 We employ  parameters of reference \cite{32, 33} as follow
  $\Lambda=602.3 MeV$.  ;
$G\Lambda^2=1.835\Lambda$ and $K\Lambda^2=12.36 $. $G$ and $K$ are
coupling strength.

The dynamical mass is calculated by
\begin{equation}\label{eq14}
m_i=m_0^i-4G<\bar{q}_iq_i>+2K<\bar{q}_jq_j><\bar{q}_kq_k>,
\end{equation}
\begin{equation}\label{eq15}
<\bar{q}_iq_i>=-\frac{3}{\pi^2}\int_{p_{Fi}}^\Lambda
p^2dp\frac{m_i^2}{\sqrt{m_i^2+p^2}},
\end{equation}

\begin{equation}\label{eq16}
p_{Fi}=(\pi^2\rho_i)^{1/3}.
\end{equation}
At this stage  we will determine EOS of quark matter in NJL model,
\begin{equation}\label{eq18}
P=\sum_{i=u,d,s}n_i\sqrt{p_{Fi}^2+m_i^2} -{\cal E},
\end{equation}
where
\begin{equation}\label{eq19}
{\cal
E}=\sum_{i=u,d,s}\frac{3}{\pi^2}\int_0^{p_{Fi}}p^2dp\sqrt{p^2+m_i^2}-(B-B_0).
\end{equation}
In  equation (\ref{eq19}), $B$ is the bag pressure \cite{13}
which is consequence of  interaction,
\begin{eqnarray}\label{eq20}
B&=&\sum_{i=u,d,s}\left[\frac{3}{\pi^2}\int_0^\Lambda
p^2dp\left(\sqrt{p^2+m_i^2} - \sqrt{p^2+{m_0^i}^2}\right)\right.\nonumber\\&&\left. -2G<\bar{q}_iq_i>^2\right]
+4K<\bar{u}u><\bar{d}d><\bar{s}s>.
\end{eqnarray}

%%%%%%%%%%%%%%%%%%%%%%%%%%%%%%%%%%%%%%%%%%%%%%%%%%%%%%%%%%%%%%%%%%%%%%%%%%%%%%%%%%%%%%%%%%%%%%%

\subsection{Mixed phase}
There is a mixed phase of quarks and hadrons within a finite rang of
density.
 In this phase, we apply the Gibss condition. According to this equilibrium condition, the
pressures and chemical potentials of both quark and hadron phases
are equal \cite{4},
\begin{equation}\label{eq21}
\mu^Q_n=\mu^H_n,
\end{equation}
\begin{equation}\label{eq22}
\mu^Q_p=\mu^H_p,
\end{equation}
where $\mu^H_n$ ($\mu^H_p$) and $\mu^Q_n$ ($\mu^Q_p$) are the
neutrons
(protons) chemical potential in nucleonic and quark part in the mixed
phase, respectively. Using above equations, we can determine the
charge density of quarks and hadrons.  $\chi $, which is the volume
fraction occupied by quarks is determined by considering the global
charge neutrality. Then baryon and total energy densities  in mixed
phase could be calculated.
\begin{equation}\label{eq23}
\chi(\frac{2}{3}\rho_u-\frac{1}{3}\rho_d-\frac{1}{3}\rho_s)+(1-\chi)\rho_p-\rho_e=0,
\end{equation}
\begin{equation}\label{eq24}
\rho_B=\chi \rho_Q+(1-\chi)\rho_H,
\end{equation}
\begin{equation}\label{eq25}
{\cal E}_{MP}=\chi{\cal E}_{QP}+(1-\chi){\cal E}_{HP}.
\end{equation}
See reference  \cite{41} for detail of calculations regarding the
EOS of mixed phase.

\begin{figure}
\includegraphics{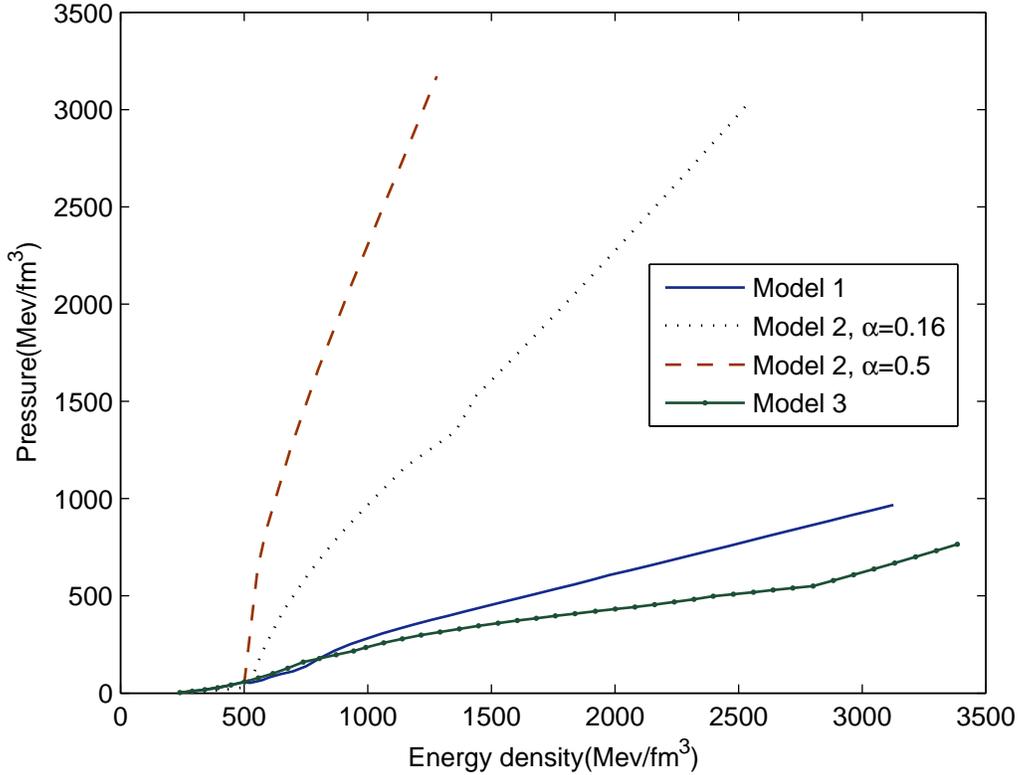}
 \caption{EOS for the hybrid neutron star. The results of our calculations in different models of
 quark matter have been plotted.}
\label{en}
\end{figure}

At this stage we can determine EOS of neutron star with a quark core
using the results of proceeding sections. We consider three
approaches for the quark matter (in the quark phase and mixed
phase); simple MIT bag model (model 1), MIT bag model by considering
the quark interaction (model 2) and NJL model (model 3). EOS results
for neutron star (corresponding to different models for the quark
matter calculations) are given in Fig. \ref {en}.
%%%%%%%%%%%%%%%%%%%%%%%%%%%%%%%%%%%%%%%%%%%%%%%%%%%%%%%%%%%%%%%%%%%%%%%%%%%%%%%%%%%%%%%%%%%%%%%%%%%%%%
%\newpage
%%%%%%%%%%%%%%%%%%%%
This figure shows that the difference between the equation of state
in model 1 and  model 3 is substantially lower than those of other
mentioned models.
%and it seems that considering the dynamical mass has no effect on equation of state.
However, we see that the equation of state of neutron star becomes stiffer
when coupling constant is higher, particulary at high densities.
This shows the importance of interaction at higher densities.

%%%%%%%%%%%%%%%%%%%%%%%%%%%%%%%%%%%%
\section{Structure properties of the hybrid Neutron Star }
\label{structure}

Using the obtained EOS in previous section, the structure of neutron
star with a quark core can be calculated. Before this work, we investigate the energy
and stability conditions for our results. For this purpose at first,
we fit a polynomial function for the equations of state in Fig.
\ref{en} as ,
\begin{equation}\label{eq26}
P=\sum_{i=1}^{7}a_{i}{\cal E}^{7-i}.
\end{equation}
 The coefficients $a_i$ have been given in Table \ref{tabcoef}.
%%%%%%%%%%%%%%%%%%%%%%%%%%%%%%%%%%%%%%%%%%%%%%%%%%%%%%%%%%%%%%%%%%%%%%%%%%%%%%%%%%%%%%%%%%%%%%%%%%%%%%
%
\begin{table}[h]
\label{tabcoef}
\begin{center}
  \caption[]{Different coefficients presented in Eq. (\ref{eq26}) for our applied models.}
\end{center}
  \begin{tabular}{|c|c|c|c|c|c|c|c|}
\hline
Model&$a_1\ (\times10^{-57})$&$a_2\ (\times10^{-40})$&$a_3\ (\times10^{-25})$&$a_4\ (\times10^{-10})$&$a_5\ (\times10^{6})$&$a_6\ (\times10^{21})$&$a_7\ (\times10^{35})$\\
\hline
  model 1  & $1.194$& $-0.2467$& $2.011$& $-8.123$& $1.656$& $-1.201$& $2.915$\\
  model 2; ${\alpha=0.16}$ & $6.487$& $-1.279$& $9.852$& $-37.23$& $7.090$& $-4.962$& $10.80$\\
  model 2; ${\alpha =0.5}$ & $9.400$& $-1.912$& $15.11$& $-58.19$& $11.20$& $-6.819$& $11.27$\\
  model 3   & $1.285$& $-0.2407$& $1.793$& $-6.630$& $1.209$& $-0.7027$& $1.321$\\
\hline
  \end{tabular}
%\end{center}
\end{table}
%%%%%%%%%%%%%%%%%%%%%%%%%%%%%%%%%%%%%%%%%%%%%%%%%%%%%%%%%%%%%%%%%%%%%%%%%%%%%%%%%%%%%%%%%%%%%%%%%%%%%
%
We use this relation to justify the energy and stability
conditions as follows.

\subsection{Energy conditions}
Energy conditions in the center of neutron star include the null
energy condition (NEC), weak energy condition (WEC), strong energy
condition (SEC) and dominant energy condition (DEC). These condition
are expressed as follow,
\begin{equation}\label{eq27}
NEC\rightarrow P_{c}+{\cal E}_{c}\geq0,
\end{equation}

\begin{equation}\label{eq28}
WEC\rightarrow P_{c}+{\cal E}_{c}\geq0\quad\&\quad     {\cal
E}_{c}\geq0,
\end{equation}

\begin{equation}\label{eq29}
SEC\rightarrow P_{c}+{\cal E}_{c}\geq0  \quad\&\quad
 3P_{c}+{\cal E}_{c}\geq0,
\end{equation}

\begin{equation}\label{eq30}
DEC\rightarrow {\cal E}_{c}>|P_{c}|,
\end{equation}
where $ {\cal E}_{c} $ is the energy density and $P_c$ is the
pressure at the center of  star. Results of the above conditions for
our equations of state are given in Table \ref{tab1}.
% by Considering figure \ref{en} and the above conditions.
%%%%%%%%%%%%%%%%%%%%%%%%%%%%%%%%%%%%%%%%%%%%%%%%%%%%%%%%%%%%%%%%%%%%%%%%%%%%%%%%%%%%%%%%%%%%%%%%%%%%%%
%\newpage
\begin{table}[h]
\label{tab1}
\begin{center}
  \caption[]{Energy conditions for hybrid neutron star for applied models.}
  \begin{tabular}{clclclclclcl}
  \hline\noalign{\smallskip}
 & NS+Quark Core & ${\cal E}_c\left({10^{14}gr/cm^3}\right)$  & $P_c\left({10^{14}gr/cm^3}\right)$ & $ NEC $ & $ WEC $ & $ SEC $ & $ DEC $ \\
 \hline\noalign{\smallskip}

 & model 1  & 25.8 & 7.79 & $\surd$ & $\surd$ & $\surd$ & $\surd$ \\
 & model 2 ; ${\alpha=0.16}$  & 18.25 & 17.8 & $\surd$ & $\surd$ & $\surd$ & $\surd$\\
 & model 2 ; ${\alpha
 =0.5}$  & 14.05 & 28.9 & $\surd$ & $\surd$ & $\surd$ & $\times$\\
 & model 3  & 22.9 & 5.56 & $\surd$ & $\surd$ & $\surd$ & $\surd$ \\
 \noalign{\smallskip}\hline
  \end{tabular}
\end{center}
\end{table}
%%%%%%%%%%%%%%%%%%%%%%%%%%%%%%%%%%%%%%%%%%%%%%%%%%%%%%%%%%%%%%%%%%%%%%%%%%%%%%%%%%%%%%%%%%%%%%%%%%%%%%%%%%%%%%%%%%%%%%%%
We have found that our equations of state satisfy mentioned energy
conditions, except the dominant energy condition for model 2 with
$\alpha=0.5$ in which the central pressure is very high with respect
to the other models.

\subsection{Stability}
\begin{figure}
\includegraphics{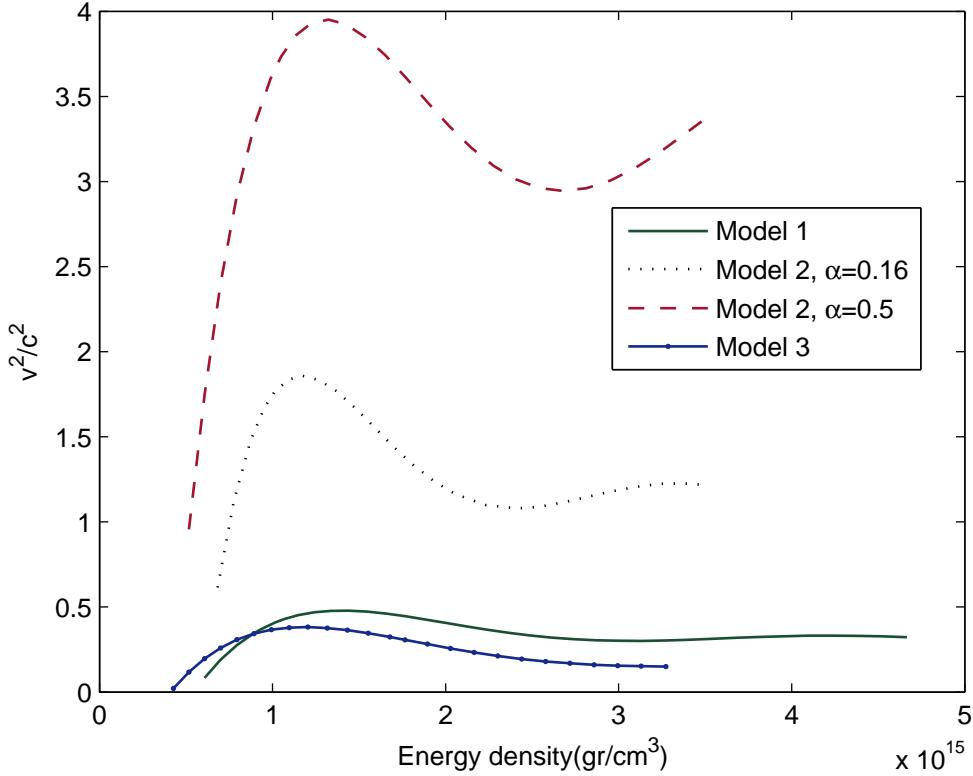}
 \caption{Sound speed versus density for different models.}
\label{vc}
\end{figure}

According to the stability condition, an equation of state is
physically acceptable when the corresponding obtained velocity of sound
($v$) be less than the light's velocity (c) \cite{35, 36}.
 Thus the stability condition is
$(0\leq v^2=(\frac{dP}{d\cal E})\leq c^2)$. By
 Using Eq. \ref {eq26}, we have computed $\frac{v^{2}}{c^{2}}$ versus
 density which has been given
 in Fig. \ref{vc}. It is evident that the stability
condition is satisfied by the our calculated EOS of neutron star
with quark core for models 1 and 3.
%%%%%%%%%%%%%%%%%%%%
%%%%%%%%%%%%%%%%%%%%%%%%%%%%%%%%%%%%%%%%%%%%%%%%%%%%%%%%%%%%%%%%%%%%%%%%%
So these two models are suitable for determining the structure of a
hybrid neutron star. The other models do not obey the stability
condition, and we can't use their equations of state in the structure
calculations.
%%%%%%%%%%%%%%%%%%%%%%%%%%%%%%%%%%%%%%%%%%%%%%%%%%%%%%%%%%%%%%%%%%%%%%%%%%%%%%%%%%%%%%%%%%%%%%%%%%%%%%%%%%%%%%%%%%%%%%%%%%%%

%At this time we calculate structure of star with two models 1 and 3.

\subsection{Properties of the neutron star with a quark core }
 We use TOV equation to calculate the structure of star. This equation is determined by Tolman-Oppenheimer-Volkoff  (TOV)
\cite{37, 38, 39}.
\begin{equation}\label{eq27}
    \frac{dP}{dr}=-\frac{G\left[{\frac{P}{c^2}+\cal E}\right]
    \left[m+\frac{4\pi r^3 P}{c^2}\right]}{r^2
    \left[1-\frac{2Gm}{rc^2}\right]},
\end{equation}

\begin{equation}\label{eq28}
    \frac{dm}{dr}=4\pi r^2{\cal E}
\end{equation}
where $P$ is the pressure and $\cal E$ is the energy density.
In our calculations for the neutron star, we consider the following
equation of state: up to the density $0.05 fm^{-3}$, we use the data
of Baym calculations  \cite{40}. For the hadron, quark, and
mixed phases, we consider our equations of state obtained in the previous
sections.
 Numerically integrating the TOV equation for a given equation of state,
 the mass and radius of the neutron star with a quark core is determined.
Our results are as follows.

\begin{figure}
\includegraphics{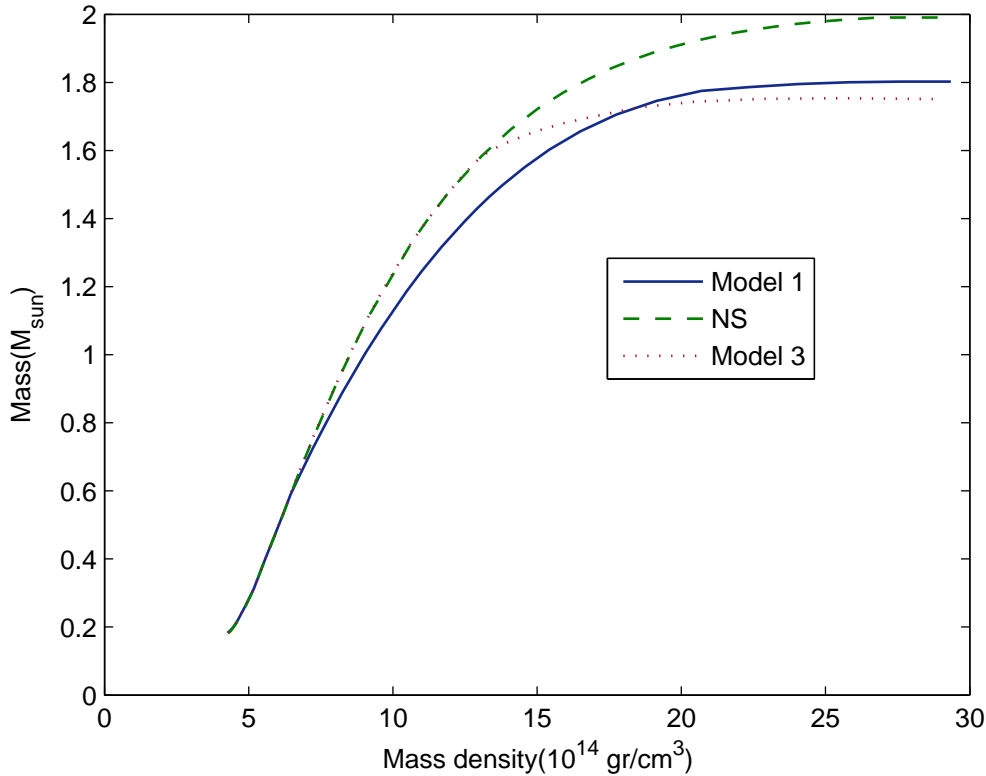}
 \caption{Gravitational mass versus the central mass density for the
 neutron star with a quark core for different models. The  results for a neutron star without quark core (NS) have been also given for comparison.}
\label{mass}
\end{figure}

\begin{figure}
\includegraphics{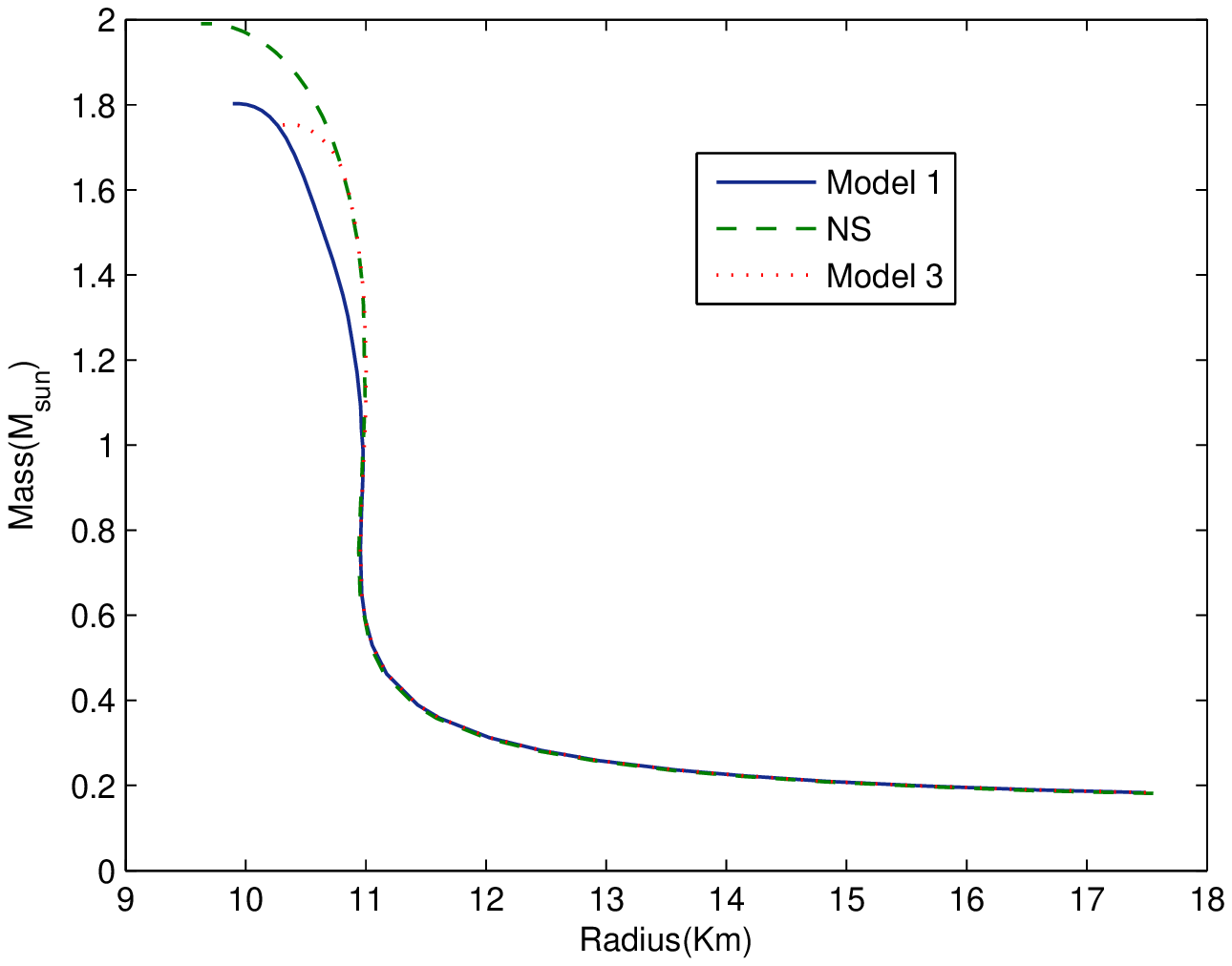}
 \caption{Mass-radius relation for the neutron star
 with a quark core for different models. The  results of neutron star without quark core (NS) have been also given for comparison.}
\label{radius}
\end{figure}

The gravitational mass of hybrid neutron star (a neutron star with a quark core) have been presented in Fig.
 \ref{mass} versus the central
mass density. The mass-radius relation has been also shown in Fig. \ref{radius}
for this star for different models. For comparison, we have also brought the results for a neutron
 star without a quark matter \cite{41} in Figs. \ref{mass} and \ref{radius}.
 %%%%%%%%%%%%%%%%%%%%%%%%%%%%%%%%%%%%%%%%%%%%%%%%%%%%%%%%%%%%%%%%%%%%%%%%%
%\newpage
%%%%%%%%%%%%%%%%%%%%%%%%%%%%%%%%%%%%%%%%%%%%%%%%%%%%%%%%%%%%%%%%%%%%%%%%%
%\newpage
%%%%%%%%%%%%%%%%%%%%%%%%%%%%%%%%%%%%%%%%%%%%%%%%%%%%%%%%%%%%%%%%%%%%%%%%%%%
Results show that in the NJL model with a
density dependent mass of quarks, the maximum mass is lower than that of
other models.
 The structure properties
of neutron star in the cases without and with a quark and  have been
given in Table \ref{tab4} for different models.
%%%%%%%%%%%%%%%%%%%%%%%%%%%%%%%%%%%%%%%%%%%%%%%%%%%%%%%%%%%%%%%%%%%%%%%%%%%%%%%%%%%%%%%%%%%%%%%%%%
%%%%%%%%%%%%%%%%%%%%%%%%%%%%%%%%%%%%%%%%%%%%%%%%%%%%%%%%%%%%%%%%%%%%%%%%%%%%%%%%%%%%%%%%%%%%%%%%%%%%%%%%%%%%%%%%%%%%%%%%
%
\begin{table}[h]
\label{tab4}
\begin{center}
  \caption{Structure properties of  neutron star without (NS) and with (NS+Q) a quark core for different models. The observational data have been also presented for comparison.}
  \begin{tabular}{clclclcl}
  \hline\noalign{\smallskip}
 & Compact object  & $ M_{max}\left(M_{\,\odot}\right) $ & $ R\left(km\right) $  \\
 \hline\noalign{\smallskip}
  & NS & 1.98 & 9.8  \\
 & NS+Q: model 1  & 1.8 & 10\\
 & NS+Q: model 3  & 1.75 & 10.4  \\
 & $4U1820-30$ \cite{42}&$1.58\pm0.6$  &  $9.1\pm0.4$  \\
 & $PSR J1903+0327$ \cite{43} & $1.667\pm0.021$  &  $9.438 km$ \\
 & $PSR J1614-2230$ \cite{44}& $1.97\pm0.04$  & $13\pm2km$  \\
 &$SAX J1748.9-2021$\cite{45} & $1.78\pm0.3$  & $8.18\pm1.62km$   \\
 & $4U1608-52$ \cite{46} & $1.74\pm0.14$&$9.3\pm1km$\\
 & $Vela X-1$ \cite{46}& $1.77\pm0.08$ &$9.56km$\\
 \noalign{\smallskip}\hline
  \end{tabular}
\end{center}
\end{table}
%
%%%%%%%%%%%%%%%%%%%%%%%%%%%%%%%%%%%%%%%%%%%%%%%%%%%%%%%%%%%%%%%%%%%%%%%%%%%%%%%%%%%%%%%%%%%%%%%%%%%%%%%%%%%%%%%%%%%%%%%%%
\begin{table}[h]
\label{tab5}
\begin{center}
  \caption{Structure properties of  neutron star with a quark core is calculated by other authors. Our results have been also given for comparison.}
  \begin{tabular}{c|c|c|c|c|c|c|c|l|}
  \hline\noalign{\smallskip}
&$ Reference $ & $ Quark\,Phase $ & $ Hadron\,Phase $ & $ Mixed\,Phase $ & $ used$ & $used $ & $M_{max}\left(M_{\,\odot}\right)$ & $R\left(km\right)$ \\
&  & $model$ & $model$ & $condition$ &$ parameter$  & $ parameter$ &  & \\
 \hline
 &present wok&MIT Bag Model &LOCV &Gibbs   &  &  & 1.8 & 10\\
 &present work& NJL Model&LOCV & Gibbs & $$ & $$ & 1.75 &10.4\\
 \hline\noalign{\smallskip}
 &  &  &  &  & $x_{\sigma}=0$ & $ G_{2}=0.006$ & 1.44 & 9.54\\
& & &  &  & $x_{\sigma}=0$ & $G_{2}=0.012$ & 1.89 & 12.55\\
 & & &  &  & $x_{\sigma}=0$ &$G_{2}=0.016$ & 2.05 & 12.66\\
 & \cite{46}&Field correlator& Relativistic&   & $x_{\sigma}=0.8$ & $G_{2}=0.006$ & 1.44 & 9.52\\
  && Method &Mean Field  & Gibbs & $x_{\sigma}=0.8$ & $G_{2}=0.012$ & 1.89 & 12.53\\
 & & & Model&  & $x_{\sigma}=0.8$ & $G_{2}=0.016$ & 2.04 & 12.40\\
 & & &  &  & $x_{\sigma}=0.6$ & $G_{2}=0.006$ & 1.43 & 9.51\\
 & & &  &  & $x_{\sigma}=0.6$ & $G_{2}=0.012$ & 1.73 & 12.00\\
 & & &  &  & $x_{\sigma}=0.6$ & $G_{2}=0.016$ & 1.8 & 11.49\\
\hline\noalign{\smallskip}
 &   &  &  &  &$\sigma=0$  &$G_{v}=0$   & 1.91 & 13.09\\
   & &  &  &  &$\sigma=0$  & $G_{v}=0.2G_{s}$ & 2.05 & 13.00\\
  &  &  &  &  &$\sigma=0$  & $G_{v}=0.4G_{s}$ & 2.13 & 12.77\\
 & \cite{47} & NJL Model & Relativistic  &  &$\sigma=10$  & $G_{v}=0$ & 1.94 & 13.3\\
&    &  & Mean Field & Gibbs &$\sigma=10$  & $G_{v}=0.2G_{s}$ & 2.08 & 13.01\\
  &  &  & Model &  &$\sigma=10$  & $G_{v}=0.4G_{s}$ & 2.15 & 12.77\\
 &   &  &  &  &$\sigma=40$  & $G_{v}=0$ & 2.00 & 13.37\\
 &  &  &  &  &$\sigma=40$  & $G_{v}=0.2G_{s}$ & 2.11 & 13.03\\
  &  &  &  &  &$\sigma=40$  & $G_{v}=0.4G_{s}$ & 2.17 & 12.67\\
\hline\noalign{\smallskip}
 &   &  &  &  &  & $\beta=0.000$ & 2.029& 11.13\\
 &  &  &  &  &  & $\beta=0.025$ & 2.003 & 11.56\\
 & \cite{48}  & MIT & Chiral Quark- & Gibbs & ESO8 & $\beta=0.050$ & 1.958 & 11.71\\
  &  & Bag Model & Meson Coupling  &  &  & $\beta=0.100$ & 1.896 & 11.63\\
  &  &  & Model &  &  & $\beta=0.150$ & 1.866 & 11.5\\
  &  &  &  &  &  & $\beta=0.200$ & 1.853 & 11.41\\
 \hline\noalign{\smallskip}
  \end{tabular}
\end{center}
\end{table}
%%%%%%%%%%%%%%%%%%%%%%%%%%%%%%%%%%%%%%%%%%%%%%%%%%%%%%%%%%%%%%%%%%%%%%%%%%%%%%%%%%%%%%%%%%%%%%%%
In table \ref{tab4}, it is seen that when we consider a quark core for the
neutron star in both considered
 models, the maximum mass decreases and radius increases. This is because the equation of state becomes softer when we consider a
 quark matter in the core of neutron star.
 We have
 obtained the lowest amount of mass in NJL model $(mass=1.75M_{\,\odot})$, but it
is nearly close to that of MIT bag model with zero coupling constant
$(mass=1.8M_{\,\odot})$. The calculated radius in NJL model is bigger than that of MIT bag model.
%
%%%%%%%%%%%%%%%%%%%%%%%%%%%%%%%%%%%%%%%%%%%%%%%%%%%%%%%%
We have also brought some observational data of neutron star
candidates in table \ref{tab4} to compare our results with these data.
 The mass of a neutron
star with a quark core which we have been determined in this paper is closed to the mass of $SAX J1748.9-2021$ \cite{46}, $4U1608-52$ \cite{47},  and $Vela X-1$ \cite{47}. Also the calculated radius of neutron star with a quark core in two models are in a good agreement with the radius of the mentioned observed compact objects.  However, the mass and radius do
not agree with the recent observational data of for pulsar $PSR
J1614-2230$ \cite{45} with $M =1.97\pm0.04M_{\,\odot}$, while our result of
neutron star without quark core
\cite{41} has a good agreement with the mass of this pulsar. But the calculated radius is smaller than the radius of pulsar $PSR
J1614-2230$.

 Also we review the works of several authors who
have researched on
 the properties of hybrid neutron star. The results of their models are
 presented briefly in table \ref{tab5}. In this table, $x_{\sigma}$ is the hyperon coupling
  constant and $G_{2}$ is one of the EOS parameters within the field
  correlated method \cite{47}.
  We can see that the results of reference \cite{47} with $x_{\sigma}=0.6$, and $G_{2}=0.012$ and $0.016$  ($M=1.73M_{\,\odot}$ and $1.8M_{\,\odot}$)
  are in a good agreement with our calculated mass.
  In table \ref{tab5}, $\sigma$ and $G_{v}$ are the surface tension
 and the vector coupling constant, respectively \cite{48}. Mass values in that paper are about $2M_{\,\odot}$, therefore they have a good agreement with $PSR J1614-2230$, while they do not  have agreement with our work.
  In table \ref{tab5}, ESCO8 is a model for determining hyperon coupling constant and $\beta$ is a parameter which is the same with
  the one in bag constant formula, $B$, in our work \cite{49}. The obtained masses in this reference is higher than our calculated result. The authors in all these papers
  have parameterized the EOS in order to get $M\cong
  2.0M_{\,\odot}$ which has been recently observed for $PSR J1614-2230$.
  Here, we can conclude that if we
want to get the mass of a hybrid star to be about the mass of pulsar
$PSR J1614-2230$ with the mass $M =1.97\pm0.04M_{\,\odot}$, we
should modify the equation of state that used in this work,
or we should use another computational method to calculate the
structure of a hybrid star.

%%%%%%%%%%%%%%%%%%%%%%%%%%%%%%%%%%%%%%%%%%%%%%%%%%%%%%%%%%%%%%%%%%%%%%%%%%%%%%%%%%%%%%%%%%%%%%%%%%%%%%%%%%%%%%%%%%%%%%%%%%%%

%Now we investigate other properties of a hybrid in the next sections.

 %%%%%%%%%%%%%%%%%%%%%%%%%%%%%%%%%%

\section{Summary and Conclusion}
Since the neutron stars are one of the compact objects with high
density, this idea raises that there is  a deconfined quark
matter in these stars. Thus, here we considered a crust of hadronic
matter, a mixed phase of quark and hadronic matters and a quark
matter in the core of neutron star. In this work, we calculated the EOS
for quark matter phase of neutron star in three models,
simple MIT bag model, MIT bag model including one gluon exchange correction with two
different coupling constants and NJL model with a density
dependent mass for quarks. For hadronic matter phase, we chose a
variational method (LOCV). For the mixed phase, the hadron-quark
phase transition was modeled by the Gibbs constructions. After
calculation of EOS, we studied the energy and stability conditions.
We found that when the interaction of quarks is given by the one gluon exchange,
the equation of state doesn't satisfy these conditions. Therefor in
this case EOS isn't suitable for calculation of structure of this
star. Using determined equation of state and solving
Tolman-Oppenheime-Volkof (TOV) equations, we computed the structure
of a neutron star with a quark core (hybrid star) in two models 1 and 3.
We saw that our result for the maximum mass of neutron star with a quark core
agrees with the observed mass for $SAX J1748.9-2021$, $4U1608-52$  and $Vela X-1$.
However, that is not in a good agreement with the recent observational data for $PSR
J1614-2230$. While, we found that our result for the maximum mass of neutron star without quark
matter is in agreement to the mass of this object. Here, it can be concluded that our equation of state
should be modified in order to get some good agreements with the new observational data.

%%%%%%%%%%%%%%%%%%%%%%%%%%%%%%%%%%%%%%%%%%%%%%%%%%%%%%%%%%%%%%%%%%%%%%%%%%%%%%%%%%%%%%%%%%%%%%%%%%
\section*{Acknowledgements}
{We wish to thank the Research Council of Islamic
Azad University, Bafgh Branch.
We also wish to thank Shiraz
University Research Council. This work has been supported
financially by the Research Institute for Astronomy and
Astrophysics of Maragha, Iran.
G. H. Bordbar wishes to thank N. Afshordi (University of Waterloo) for his useful comments and discussions during this work.
G. H. Bordbar also wishes to thank Physics Department of University of Waterloo
for the great hospitality during his sabbatical.
}

%%%%%%%%%%%%%%%%%%%%%%%%%%%%%%%%%%%%%%%%%%%%%%%%%%%%%%%%%%%%%%%%%%%

%%%%%%%%%%%%%%%%%%%%%%%%%%%%%%%%%%%%%%%%%%%%%%%%%%%%%%%%%%%%%%%%%%%%%%%%%%%%%%%%%%%%%%%%%%%%%%%%%%%%%%

%%%%%%%%%%%%%%%%%%%%%%%%%%%%%%%%%%%%%%%%%%%%%%%%%%%%%%%%%%%%%%%%%%%%%%%%%
\end{document}